


\hsize 6.5truein
\vsize 9.0truein
\hoffset 0.2truein
\voffset 0.06truein

\def\singlespace{\baselineskip 12pt}

\def\medspace{\baselineskip 18pt}

\def\doublespace{\baselineskip 24pt}
\def\bk{\hfill\break}
\def\endpage{\vfill\eject}

\def\mc{m_\chi^2}
\def\echi{E_{\delta\chi}}
\def\ephi{E_{\delta\phi}}

\def\sqr#1#2{{\vcenter{\vbox{\hrule height.#2pt
       \hbox{\vrule width.#2pt height#1pt \kern#1pt
          \vrule width.#2pt}
       \hrule height.#2pt}}}}

\def\phib{\bar\phi}

\def\kmax{k_{\rm max}}

\def\phik{\phi_{\rm kink}}
\def\dplus{\delta^+}
\def\dminus{\delta^-}

\def\veff{V_{\rm eff}}

\def\m#1{m_{#1}}
\def\phim{\phi_M}

\def\cutoff{\Lambda}
\def\gl{{g \over \lambda}}

\doublespace
\PhysRevtrue
\nopubblock

\titlepage
\singlespace
\rightline{\hfill FERMILAB-Pub-92/281-A}
\rightline{\hfill UCLA/92/TEP/37}
\rightline{\hfill October 1992}
\medspace

\title {HOW \ EFFECTIVE \ IS \ THE \ EFFECTIVE \ POTENTIAL ?}
\bk

\author {Scott Dodelson$^{1,}$\footnote{\sharp}{E-mail address:
Dodelson@fnal.fnal.gov} and Ben-Ami
Gradwohl$^{1,2,}$\footnote{\natural}{E-mail address:
Gradwohl@uclahep.bitnet}}

\address {$^1$NASA/Fermilab Astrophysics Center\break
          Fermi National Accelerator Laboratory\break
          P.O. Box 500, Batavia, IL 60510 \break
                    \break
          $^2$Department of Physics \break
          University of California Los Angeles\break
          Los Angeles, CA 90024}

\baselineskip 13pt
\vskip 0.4in
\centerline{ABSTRACT}
\vskip 0.25in
Motivated by bubble nucleation in first order phase transitions,
we question the validity of
the effective potential for inhomogeneous configurations. In an attempt
to get some insight into the importance of derivative terms, we analyze a
simple model, a kink in (1+1) dimensions and zero temperature.
We evaluate the energy shift from the quantum fluctuations about the
non-uniform background ({\it i.e.}, the effective action) and compare it
to the energy from the effective potential. Our results clearly show
that for inhomogeneous configurations it may be inadequate to
omit derivative terms and confine oneself to the effective potential.
We then couple the kink field to an additional scalar field and perform
the same comparison. The addition of the second field allows us to vary
the mass of the fluctuations and their coupling to the underlying kink.
If the mass of the second field is large, it does not feel the inhomogeneities
in the kink field and consequently does not give rise to important derivative
corrections in the effective action. In contrast, if the mass is small,
derivative terms are significant and the effective potential fails.
In the latter regime we can, however, rely
on the Born approximation to calculate the effective action.

\endpage
\doublespace

\FIG\FKINKCUT{The effect of the kink on the frequencies and on the
cut-off. Open squares represent wavevectors $k$ in the absence of the kink.
In this artificial example, the momentum cut-off $k=\Lambda$ corresponds
to including only those states with $n\le 100$. In the presence of the kink
(crosses), the same value of $n$ has a different wavenumber. In particular,
for $n=100$, $k$ is less than $\Lambda$. Blindly imposing the ultraviolet
cut-off $k\le \Lambda$ in the presence of the kink leads to the counting
of an extra state, the one with $n=101$.}

\FIG\FSHIFT{Even (solid line) and odd (dashed line) parity phase shifts
of the $\chi$ continuum, as a function of wavenumber {\it q} for
fixed $g/\lambda$. At large $q$, the phase shift goes to zero; only at low $q$
does the presence of the kink lead to a significant phase shift.
Larger values of
$g/\lambda$ correspond to deeper potentials and hence larger phase shifts.}

\FIG\FSHIFTZERO{Even (solid line) and odd (dashed line) phase
shifts for the continuous $\chi$ states at zero momentum,
as a function of $g/\lambda$.}

\FIG\FEN{A comparison of the energy shift from the $\chi$-fluctuations,
as a function of $g / \lambda$,
calculated within the effective action (solid line) and the effective potential
(short dashed line) approaches. For large $g / \lambda$ the derivative terms
are
unimportant and the effective potential becomes a valid approximation to the
full action. For small $g / \lambda$, however, the effective potential
is clearly inadequate for calculating the energy of an inhomogeneous
configuration. We also present the energy shift, computed with in
the Born approximation (dotted line).}

\FIG\FLOOP{Feynman diagram leading to terms $\phi^4, \phi^2 ( \partial_
\mu \phi )^2, \ldots$ in the effective action.
Dotted line represents the internal $\chi$ propagator.}


\REF\QCD{ See, {\it e.g.}, F. Wilczek, preprint
IASSNS-HEP-92/23 (1992).}

\REF\BBN{J. H. Applegate and C. J. Hogan, {\it Phys. Rev. D}
{\bf 31} (1985) 3107;\bk
J. Applegate, C. Hogan, and R. J. Scherrer, {\it
Phys. Rev. D} {\bf 35} (1987) 1151.}

\REF\ELECTRO{D. A. Kirzhnits and A. D. Linde, {\it Ann. Phys.} {\bf 101}
(1976) 195.}

\REF\ANDERSON{G. Anderson and L. Hall, {\it Phys. Rev. D} {\bf 45}
(1992) 2685.}

\REF\CARRINGTON{
M. E. Carrington, {\it Phys. Rev. D} {\bf 45} (1992) 2933;\bk
M. Dine, R. G. Leigh, P. Huet, A. Linde, and D. Linde, Stanford University
preprints SU-ITP-92-6 and SU-ITP-92-7;\bk
D. E. Brahm and S. D. H. Hsu,
Caltech preprints CALT-68-1705 and CALT-68-1762;\bk
C. G. Boyd, D. E. Brahm, and S. D. H. Hsu, Caltech preprint CALT-68-1795;\bk
M. Gleiser and E. W. Kolb, FNAL preprint FERMILAB-Pub-92/222-A
(1992).}

\REF\ROCKY{
M. Gleiser and E. W. Kolb, {\it Phys. Rev. Lett.} {\bf 69} (1992) 1304.}

\REF\BAS{M. E. Shaposhnikov, {\it JETP Lett.} {\bf 44} (1986) 465;
{\it Nucl. Phys. B} {\bf 287} (1987) 757; {\it Nucl. Phys. B} {\bf 299}
(1988) 797;\bk
L. McLerran, {\it Phys. Rev. Lett.} {\bf 62} (1989) 1075;\bk
A. I. Bochkarev, S. Yu. Khlebnikov, and M. E. Shaposhnikov,
{\it Nucl. Phys. B} {\bf 329} (1990) 490;\bk
N. Turok and P. Zadrozny, {\it Phys. Rev. Lett.} {\bf 65} (1990) 2331;
{\it Nucl. Phys. B} {\bf  358} (1991) 471;\bk
L. McLerran, M. E. Shaposhnikov, N. Turok, and M. Voloshin,
{\it Phys. Lett. B} {\bf 256} (1991) 451;\bk
M. Dine, P. Huet, R. Singleton, and L. Susskind, {\it Phys. Lett. B}
{\bf 257} (1991) 351 ;\bk
A. Cohen, D. B. Kaplan, and A. E. Nelson, {\it
Nucl. Phys. B} {\bf 349} (1991) 727; {\it Phys. Lett. B} {\bf 263}
(1991) 86;\bk
B. H. Liu, L. McLerran, and N. Turok, Minnesota preprint TPI-MINN-92/18-T.}

\REF\COLEMAN{S. Coleman, {\it Phys. Rev. D} {\bf 15}, 2929 (1977);
\bk
C. Callan and S. Coleman, {\it Phys. Rev. D} {\bf 16}, 1762 (1977).}

\REF\ASPECTS{
S. Coleman, {\it Aspects of Symmetry} (Cambridge University Press, 1985).}

\REF\BRANDEN{R.H. Brandenberger \journal Rev. Mod. Phys. & 57 (85) 1.}

\REF\EXPANSION{R. Jackiw \journal Phys. Rev. & D9 (74) 1686;\bk
J. Iliopoulos, C. Itzykson, and A. Martin \journal Rev. Mod. Phys.
& 47 (75) 165;\bk
L.-H. Chan, {\it Phys. Rev. Lett.} {\bf 54} (1985) 1222; {\it Phys.
Rev. Lett.} {\bf 56} (1986) 404 (E);\bk
O. Cheyette \journal Phys. Rev. Lett. &55 (85) 2394;\bk
C.M. Fraser \journal Z. Phys. & C28 (85) 101;\bk
I.J.R. Aitchison and C.M. Fraser \journal Phys. Rev. & D32 (85)
2190;\bk
I. Moss, D. Toms, and A. Wright \journal Phys. Rev. &D46 (92) 1671.}

\REF\DASHEN{R. F. Dashen, B. Hasslacher, and A. Neveu, {\it Phys. Rev. D}
{\bf 10} (1974) 4130. }

\REF\CAMARGO{A.F. de Camargo and R.C. Shellard \journal Phys. Rev.
&D29 (84) 1147.}

\REF\BAACKE{J. Baacke \journal Z. Phys. & C47 (90) 263;\bk
J. Baacke, preprint DO-TH-90/5 (1990).}

\REF\NUMERICAL{M. Li, R.J. Perry, and L. Wilets \journal Phys. Rev.
&D36 (87) 596; \bk
D.A. Wasson and S.E. Koonin \journal Phys. Rev. &D43 (91) 3400.}

\REF\RAJA{R. Rajaraman, {\it Solitons and Instantons} (North-Holland, 1987),
chap.~5.}

\REF\MORSE{P.M. Morse and H. Feshbach, {\it Methods of Theoretical Physics},
part II, (McGraw-Hill Book Company, 1953), section~12.3.}

\REF\LEVINSON{N. Levinson, {\it K. Dan. Vidensk. Selsk. Mat. Fys. Medd.}
{\bf 25} (1949), 710; for a review see, {\it e.g.}, M.L. Goldberger and K.M.
Watson, {\it Collision Theory} (Wiley, New York, 1964).}

\REF\BERG{B. Berg, M. Karowski, W. R. Theis, and H.J. Thun \journal
Phys. Rev. & D17 (78) 1172.}

\REF\JACKIW{R. Jackiw and G. Woo \journal
Phys. Rev. & D12 (75) 1643.}

\REF\BUTKOV{E. Butkov, {\it Mathematical Physics} (Addison Wesley, 1968),
p.631 ff.}


\chapter{Introduction}

Phase transitions have played a crucial role in the early evolution
of the universe. From a possible inflationary epoch at the GUT scale, through
the electroweak and quark-hadron transitions,
phase transitions took a major
part in shaping our universe. The significance and observable consequences of a
specific transition depend on its detailed nature,
and in particular, on whether it was
first or second order. To mention only two examples in this connection:
If the QCD phase transition~[\QCD] was first order (which at
present seems unlikely), Big Bang nucleosynthesis may have taken place
in a very inhomogeneous setting [\BBN] and the elemental
abundances
could differ significantly from the predictions of the standard scenario.
If the electroweak phase transition was first
order~[\ELECTRO]\footnote{\dagger}{See ref. \ANDERSON\
for a clear, analytic treatment which appears to
be justified by more recent treatments accounting for
higher order effects [\CARRINGTON]. For a slightly different point of view see
[\ROCKY].}, the resulting
non-equilibrium period may have generated the observed baryon asymmetry
in the universe~[\BAS].

In a second order phase transition, the field evolves smoothly from its high
temperature `false' vacuum state to its low temperature `true' vacuum,
whereas in
a first order transition the field is trapped in a local minimum.
In the latter case, the transition eventually
proceeds (at zero temperature) via tunneling through the energy barrier,
which separates the true from the false vacuum. (At
finite temperature, thermal fluctuations `push' the field over the
barrier in {\it free} energy.) This corresponds, in the overall picture, to
nucleation of true vacuum bubbles in the surrounding false vacuum sea.
The bubbles expand and coalesce, thereby completing the transition. In
general, first order phase transitions are associated with supercooling,
out-of-equilibrium processes, and the formation of shock waves in front of the
expanding bubbles. The typical time scale for the first two effects is given by
the inverse of the bubble nucleation rate or decay rate of the false
vacuum. To understand a given phase transition, we clearly have to
calculate the nucleation rate accurately.

Coleman [\COLEMAN] derived the following expression
for the tunneling rate per unit volume per unit time\footnote{\ddagger}
{For reviews, see refs. [\ASPECTS] and [\BRANDEN].},
$$ \Gamma = \left( {S(\phib)\over 2\pi \hbar } \right)^2
\bigg \vert { {\rm det} (-\partial_\mu\partial_\mu + V''(\phi_f) ) \over
{\rm det}' (-\partial_\mu\partial_\mu + V''(\phib)) }
\bigg \vert^{1/2}    e^{-S(\phib)/\hbar } \ , \eqn\EQONEO $$
where $\phi_f$ is the value of the field in
the false vacuum; the Euclidean index $\mu$ runs from $0$ to $D$, where
$D$ is the number of spatial dimensions; $S$ is
the Euclidean action, a functional of $\phi$; $V$ is the potential;
and $\phib$ is a non-trivial solution to the Euclidean equations of motion
subject to the boundary conditions: $\lim_{\vert\vec x\vert,\tau \rightarrow
\infty} \phib(\tau,\vec x) = \phi_f$. The prime on the determinant in the
denominator indicates that we exclude the (four) zero modes corresponding to
translations. Eq. \EQONEO\ tells us that the leading contribution
to the imaginary part of the energy, or the decay rate, comes from the
virtual configuration $\phib(\vec x,\tau)$. The ratio of the determinants
represents quantum fluctuations about this configuration.

Our interest in this work is in the quantum corrections, represented by
the determinant in eq. \EQONEO. Nominally one expects
corrections from the determinant to be small, so it is often sufficient
to simply evaluate the argument of the exponential and estimate
the determinant on dimensional grounds.
Nonetheless, it is precisely these quantum corrections which sometimes
transform a second order phase transition to a first order one, or
{\it vice versa}. For example, present consensus [\ANDERSON,\CARRINGTON]
is that the electroweak
phase transition is first order because of quantum corrections
from gauge bosons.

A standard method for estimating the quantum corrections is to first
integrate out quantum fluctuations about a constant background.
This gives an effective potential for $\phi$ which is then used
in the equations of motion determining $\phib(\vec x,\tau)$.
This approach is clearly inconsistent, as the configuration $\phib$
is not constant; one should really integrate out fluctuations
about a general {\it inhomogeneous} configuration.
Put differently, one should evaluate the effective {\it action},
including all derivative terms, not just an effective potential, which
drops these terms \footnote{\natural}{In
the context of the electroweak phase transition, several groups
have recently estimated the ``derivative'' corrections to the tunnelling
rate, see Dine, {\it et al.} and Brahm and Hsu in ref.~[\CARRINGTON].
There is a wealth of literature on the derivative terms in the effective
action. See ref.~[\EXPANSION] for a partial listing.}.
In the present work we question the
usefulness of the effective potential for inhomogeneous configurations.
Specifically we ask: How accurately can the effective action be
approximated by the effective potential, how important are
derivative terms?

In section~2 we briefly review the standard technique for calculating
decay rates and formalize the above questions.
If we eventually want to make contact with cosmological phase
transitions, we will have to consider a bubble at finite temperature in
a (3+1) dimensions.
As we are trying to get as much insight
into the problem as possible, we will
restrict ourselves to a very simple  system: a kink at
zero temperature in (1+1) dimensions.
Unlike a nucleating bubble, a kink is, of course, stable but
evaluating the quantum
corrections to its energy is analogous to the problem of interest.
(After all, a decay rate is really (the imaginary part of) an energy.)
The advantage of this simple system lies in the fact that the
determinant can be evaluated exactly [\DASHEN]; that is, it is
straightforward to compute all the derivative terms in the effective
action. This gives us a mean of correctly checking the effective
potential technique. In section~3 we therefore
consider the fluctuations $\delta \phi$ in the background of
a kink $\phib$ and see
how the resulting
change in energy compares with the effective potential approach. In
section~4 we couple a second scalar field to the kink field
and again see how the two approaches compare. At the end of section $4$
we introduce a new approximation, which succeeds in the regime
where the effective potential fails.
Section~5 contains our conclusions. We reserve
technical details to two appendices.

\chapter{Quantum Corrections to an Inhomogeneous Configuration}

Consider a field in an unstable configuration
$\phi_f$ with energy $E_f$. For definiteness we take $\phi$ to be a scalar
field.
In complete analogy to quantum mechanics, the decay rate is given
by~[\COLEMAN,\BRANDEN]
$$\eqalign{\Gamma \ &= \ - {2 \over \hbar} \ {\rm Im} E_f \cr
&= \ -2 \ {\rm Im} \lim_{T \rightarrow \infty} {1 \over T} \ln \ \langle \phi_f
\vert e^{-HT/\hbar} \vert \phi_f \rangle \ , } \eqn\RATEBASIC$$
where {\it H} is the Hamiltonian. We can rewrite the matrix element
in \RATEBASIC\ in terms of a functional integral,
$$\langle \phi_f \vert e^{-HT/\hbar} \vert \phi_f \rangle \ =
\ N \int {\cal D} \phi \ e ^{-S(\phi) / \hbar} \ , \eqn\RATEFUNC$$
with $\phi$ subject to the condition
$\phi(T/2)=\phi(-T/2)=\phi_f$; $S$ denotes the
Euclidean action. The standard procedure is to evaluate the functional integral
in the WKB approximation, or equivalently to one-loop.
Expanding $S(\phi)$ about a solution to the equations of motion,
$\bar \phi$,
and keeping only terms quadratic in the fluctuations $\delta \phi = \phi -
\bar \phi$, we obtain
$$\eqalign{
N \int {\cal D} \phi \ e ^{- S(\phi) / \hbar} \ \simeq & \ N
e^{- S(\bar \phi)/\hbar} \big [ \det \big ( -\partial_\mu \partial_\mu
+V''(\bar \phi) \big ) \big ] ^{-1/2} \cr
&\equiv \exp \bigg [ { - S_{\rm eff}(\bar\phi)/\hbar \bigg ] },\cr} \eqn\WKB $$
where $\bar \phi$ (the ``bounce") satisfies the boundary conditions
$\lim _{\tau \rightarrow \pm \infty} = \phi _f$ and
$\lim _{\vert \vec x \vert \rightarrow \infty} = \phi _f$, and
$S_{\rm eff}$ is the effective action which includes all one-loop quantum
corrections. We can  determine the coefficient $N$ in eq. \RATEFUNC\
by performing a dilute-gas approximation, based on the bounce
solution~[\BRANDEN]. This, together
with eqs.~\RATEBASIC\ and \WKB, \ then leads to the
final expression for the decay rate given in eq. \EQONEO.

The two different expressions on the right hand side of eq. \WKB\
represent two alternative methods for calculating the quantum corrections.
If the effective action approach is chosen, it is often convenient to
expand $S_{\rm eff}$ about a
{\it constant} $\phi$, {\it i.e.}, in powers of momentum about a point with
zero external momenta. In position space this reads
$$S_{\rm eff}(\phi) \ = \int d^4x \ [ - V_{\rm eff}(\phi)+{1 \over 2} Z(\phi)
\ \partial_\mu \phi \partial ^\mu \phi +{\cal O} ((\partial_\mu
\phi)^4)] \ . \eqn\EFFACTION$$
For constant $\phi$
only the first term survives and the effective action is entirely
given by the effective potential. Although different techniques for
computing the additional derivative corrections have
been designed~[\EXPANSION], it is computationally elaborate to go
beyond the leading terms. Alternatively we can evaluate the
quantum corrections in the form of the determinant, eq.~\WKB,\
and thereby include {\it all} derivative terms.
However, we only know of one specific higher dimensional system (and
this system is non-renormalizable) where
the determinant has been found analytically [\CAMARGO]. In every other case
one has to rely on numerical methods [\BAACKE]. This is also true for
most lower dimensional systems~[\NUMERICAL].
One of the rare exceptions is the
kink in (1+1) dimensions, where the determinant can be evaluated analytically
{}~[\DASHEN].

We therefore use the kink to probe the effective potential
approximation. It is our goal to give a clear expos\'e on the evaluation
of the determinant about an {\it inhomogeneous} background
and compare it to the estimates of the
effective potential. Although our motivation stems from the rate of
bubble nucleation, and the bubble is unstable while the kink is stable,
the difference between stability and instability is irrelevant.
(The only difference is that we do not expect a negative
eigenvalue, which would signal instability.)
The procedure outlined above can equally well be used to
calculate the energy of the kink to one-loop.

Before proceeding to the calculation, two points are in order. First,
if other fields coupled to $\phi$ are added to the problem, then the result
$\WKB$ is easily extended to
$$ \eqalign{
\int {\cal D} \phi {\cal D} \chi \ e ^{- S(\phi,\chi) / \hbar}
&\ \simeq
e^{- S(\bar \phi)/\hbar} \bigg [ \det \bigg ( -\partial_\mu \partial_\mu
+{d^2V\over d\phi^2}\bigg \vert_{\phi=\bar \phi;\chi=0} \bigg ) \bigg ] ^{-1/2}
\cr
&\ \ \ \times \bigg [ \det \bigg ( -\partial_\mu \partial_\mu
+{d^2V\over d\chi^2}\bigg \vert_{\phi=\bar \phi;\chi=0} \bigg ) \bigg ] ^{-1/2}
\ . \cr} \eqn\TWOFIELD.$$
Second, it is perhaps more convenient to rewrite the determinant
as\footnote{\sharp}{To derive this equation we apply
$\det O = \exp({\rm Tr} \ln O)$ and integrate over the time-\bk
component of the wavenumber.}
$$ \bigg [ \det \bigg ( -\partial_\mu \partial_\mu
+{d^2V\over d\phi^2}\bigg \vert_{\phi=\bar \phi} \bigg ) \bigg ] ^{-1/2}
= \exp \bigg [ -{T\over2} \sum_n \omega_n \bigg ] ,\eqn\SUMFREQ $$
where the sum is over the eigenfrequencies of the time-independent
Schr\"odinger equation
$$ \bigg [ - \partial_i\partial_i+ {d^2V\over d\phi^2} \bigg ] \psi_n(x)
= \omega_n^2 \psi_n(x). \eqn\FREQEQ $$
Here the index $i$ runs over the number of spatial dimensions; {\it T}
has dimensions of time.

\chapter{Quantum fluctuations about a kink --- the single scalar field case}
\section{Effective Action}

We focus on the classic example of
a quantized soliton, first analysed by Dashen, Hasslacher
and Neveu~[\DASHEN] (henceforth DHN; for a review, see also ref.~[\RAJA]).
It involves the following Lagrangian in $1+1$ dimensions:
$$ {\cal L} \ = \ {1\over 2} \partial_\mu\phi \partial^\mu\phi
        - V_0(\phi) \ ;\ \ \
V_0(\phi) \ = \  {\lambda\over 4} \bigg (\phi^2 - {m_0^2 \over
\lambda}\bigg)^2 \ .
\eqn\LAGONE$$
The subscript on the mass $\m0$ will be
used to differentiate the bare mass from
the physical, renormalized value, $m$. (According to our renormalization
scheme (see Appendix~A) there is no difference between the bare
and the physical
coupling constant $\lambda$.)
The subscript on $V_0$ specifies that this is the zero order
or tree level potential.
There are two degenerate vacuum states, $\phi= {\rm constant} =
\pm \m0/\sqrt{\lambda}$.
The kink is an inhomogeneous field configuration
which interpolates between these two vacua at $x=\pm\infty$.
Specifically,
$$\phik (x) = { \m0\over\sqrt{\lambda}} \tanh \bigg
( {\m0 x\over\sqrt{2} } \bigg ) \ . \eqn\PKINK $$
For topological reasons, the kink is completely stable against decay
to the vacuum.

The classical energy of the kink is
$$\eqalign{E_{cl} \ &= \ \int_{-\infty}^\infty dx \bigg [ \ {1\over 2}
\bigg ( {d \phik(x) \over dx } \bigg ) ^2 \ + \ V_0 \big ( \phik(x) \big ) \
\bigg ] \cr
&= \ {2\sqrt{2}\over 3} \ {\m0^3\over \lambda} \ . } \eqn\ECLKINK$$
Within our renormalization scheme, when the bare parameter $m_0$
is reexpressed in terms of the physical mass $m$ and the ultraviolet
cut-off $\Lambda$, there are no ``finite'' corrections to $E_{cl}$.
The only new term diverges as $\Lambda$ goes to infinity, but is
eventually cancelled by an appropriate counterterm, arising from
quantum fluctuations.

We now calculate the energy shift due to the quantum
fluctuations about the zero order kink field.  To one-loop order this
corresponds to evaluating the determinant
in eq.~\WKB.\ As in eq.~\SUMFREQ,
the quantum correction to the energy is half the sum of all eigenvalues,
$\omega_n$, of the following Schr\"odinger equation:
$$ \bigg [\ - {d^2\over dx^2} + {\partial^2 V_0 \over \partial \phi
^2} \big (
\phik(x) \big ) \ \bigg ] \psi_n(x) \
= \ \omega_n ^2 \psi_n(x) . \eqn\SCHR$$

As shown by DHN, eq. \SCHR\ has two bound (discrete) states with
eigenfrequencies $0$ (the `zero mode') and
$\sqrt{3/2} m$\footnote{\dagger}{In
calculating the corrections of order $\hbar$, we can replace
$\m0$ by $m$;
any error is of ${\cal O}(\hbar^2)$.}.
In addition there is a continuous spectrum of states with
frequencies $\omega_{\rm cont}(k_n) = \sqrt{k_n^2 + 2 m^2}$.
These states are very similar to the corresponding plane waves in the
absence of the kink, with one significant difference: they are shifted in
phase by an amount $\delta(k_n)$ relative to each other.
DHN were able to derive an analytic form for the phase shift,
$$\delta(k) = 2\pi - 2\tan^{-1}\left( \sqrt{2} k/m \right)
- 2\tan^{-1}\left( \sqrt{2} k/2m \right) . \eqn\SHIFTDHS$$
To determine the allowed $k_n$, we
quantize the wavefunction in a box of length {\it L} (taken to $\infty$ at
the end of the calculation) and impose periodic boundary conditions.
Then the allowed values of $k_n$ satisfy
$$k_n = {2\pi n\over L} - {\delta(k_n)\over L} \ . \eqn\BOUNDARY$$
Eq. \BOUNDARY\ tells us that
the wavenumber, and therefore the frequency, associated with a
given mode $n$ is modified in the presence of a kink. This leads to a
change in the ultraviolet cut-off needed to regulate the theory.
Let us call the ultraviolet cut-off in the vacuum
$\Lambda$. In the vacuum the phase shift is of course zero, so eq.
\BOUNDARY\ shows that imposing the ultraviolet cut-off is equivalent
to including only those states with
$n\le n_{\rm max} \equiv L\Lambda/2\pi$.
In the presence of the kink we must count the exact same states,
{\it i.e.}, only those with $n \le n_{\rm max}$. Therefore, denoting
the ultraviolet cut-off in this case as $\kmax$, we see from
eq.~\BOUNDARY\ that
$$\eqalign{
\kmax \ &\equiv \ {2\pi n_{\rm max} \over L} - {\delta(\kmax)\over L}
\ = \ \cutoff - {\delta(\kmax)\over L} \cr
&\simeq \Lambda - {6m\over\sqrt{2}\Lambda L}\cr} \ , \eqn\CUTOFF$$
where the last line results from applying $\tan^{-1} x
\simeq (\pi/2) - (1/x)$ to eq.~\SHIFTDHS.\
The effect of the kink on the wavenumbers and on the cut-off is illustrated
in fig.~\FKINKCUT.\

Having catalogued the eigenfrequencies of eq.~\SCHR,\
we need only
sum them up to find the total energy shift. We turn the sum over the continuous
frequencies into an integral by using the differential form of
eq.~\BOUNDARY,\ $dn = (dk L + d\delta)/2\pi$. Consequently,
the correction to the energy due to quantum fluctuations of $\phi$ is
$$ \eqalign{\ephi\ &= \ {\hbar \over 2} \sum \omega_{\rm discrete}\ +
\ {\hbar \over 2} \sum \omega_{\rm cont}\cr
\ &= \ {\hbar\over 2} \sum \omega_{\rm discrete}\
+ \ {\hbar\over 2}
\int_{-\kmax}^{+\kmax} {dk\over 2\pi} \ \bigg ( L + {d\delta\over dk} \bigg )
\ \omega_{\rm cont}\ . } \eqn\EONEPHI$$

Eq.~\EONEPHI\ is a general expression for the quantum corrections about
a background field. How involved is this computation in general?
The continuum frequencies are completely equivalent to those
in the absence of the background field.
Thus, the only parts requiring detailed knowledge of the solution to
the Schr\"odinger equation are the frequencies of the discrete states and the
phase shifts. If we go back momentarily to the problem
which motivated us, that of quantum corrections to tunneling rates in
$3+1$ dimensions,
we cannot generally expect to solve the corresponding Schr\"odinger equation
analytically. We anticipate, though, that if there are no bound
states and if the phase shift can easily be
approximated, then the quantum corrections will be relatively easy
to evaluate. In fact, in the next section we argue that these two
conditions -- no bound states and phase shift easy to approximate --
go hand in hand. This is particularly interesting, since it is precisely
this regime where
the derivative terms are important and the effective potential a
poor approximation to the full action.

Returning to the problem at hand, we can insert the phase shift
\SHIFTDHS\ into eq.~\EONEPHI\ and evaluate the integral over the
continuous eigenvalues,
$$\eqalign{
{\hbar \over 2}
\int_{-\kmax}^{+\kmax} {dk\over 2\pi} \ \bigg ( L + {d\delta\over dk} \bigg )
\ \omega_{\rm cont} &=
\ \hbar {L \over 4 \pi}
\ \bigg [ \ \kmax^2 + m^2 \bigg ( 1 +
\ln {2 \kmax^2\over m^2 } \bigg ) \ \bigg ] \cr
&\ \ \ - \
\hbar m \bigg [ \ {3 \over 2 \sqrt {2} \pi} \
\ln {2 \kmax^2 \over m^2 }  + {1 \over \sqrt{6}} \ \bigg ]
\ .\cr } \eqn\CONTSUM$$
After inserting $k_{max}$ from eq.~\CUTOFF\ and including the bound
state energy, we obtain
$$\eqalign{
\ephi & = \hbar m\left[ {1\over 2\sqrt{6}} - {3\over \sqrt{2}\pi} \right]\cr
& + \hbar {L\over 4\pi} \left[ \ \Lambda^2 +
m^2 \bigg ( \ 1 + \ln{2\Lambda^2\over m^2} \bigg ) \ \right]
{}~-~ \hbar m \ {3\over 2 \sqrt{2} \pi}
\ \ln{2\Lambda^2\over m^2} \ . \cr} \eqn\EONEFIN $$
The second line here consists of terms that go to infinity when $\Lambda,L
\rightarrow \infty$. These infinities, though, are exactly cancelled
by the infinities in $E_{cl}$, when we express
the bare mass in terms of the physical one and introduce the induced
``cosmological constant". Adding up all these contributions, the total
kink energy then becomes
$$\eqalign{
E^{action}_{\rm kink} \ &= \
{2 \sqrt{2} \over 3} {m^3 \over \lambda}
+ \ \hbar m \
\bigg ( {1 \over 2\sqrt{6}}
- {3\over \sqrt{2} \pi} \bigg ) \ + \ {\cal O}(\hbar^2) \cr
&\simeq \ {2 \sqrt{2} \over 3} {m^3 \over \lambda} \
\bigg [ \ 1 \ - \ 0.4997 \ {\hbar \lambda \over m^2}
\ + \ {\cal O}(\hbar^2) \ \bigg]\ , \cr}
\eqn\ENERACTION$$
where the superscript {\it action} indicates that the effective
action was used, {\it i.e.}, that the quantum fluctuations were
evaluated about the true inhomogeneous background field.

\section{Effective Potential Method}

Let us now recalculate the shift in the kink energy within the
effective potential approach. The recipe is simple:
First evaluate the effective potential ({\it i.e.}, integrate out the
$\phi$ fluctuations about a {\it homogeneous} background), then solve
the equations of motion for the kink with this ``improved'' potential,
and finally derive the energy of this modified kink.
To lowest order in the coupling constant (or $\hbar$)
there is, however, a shortcut to this procedure.
{}From basic quantum mechanics we recall that the first order change in the
energy due to a perturbation $V_1$ is simply
$\langle\psi\vert V_1 \vert\psi\rangle$,
where $\psi$ is the zero order wave function. Consequently, we only have to
separate out the perturbation of order $\hbar$ due to fluctuations
of $\phi$, $V_1^{\delta\phi}(\phi)$,
from the effective potential,
$$V_{\rm eff} \ = \ V_0 (m,\lambda) \ + \ V_1^{\delta\phi} \ . \eqn\EFFPOT$$
Note that $V_0 (m,\lambda)$ is in terms of the physical mass,
$V_0 = (\lambda/4)(\phi^2-(m^2/\lambda))^2$, so it is finitie.
In $V_1^{\delta\phi}$ we can
replace the bare mass by the physical one, only making an
error of ${\cal O}(\hbar^2)$.
According to eq.~(A9) of Appendix~A,
$$V_1^{\delta\phi} \
= \ {3 \hbar \over 8 \pi} \bigg (\lambda\phi^2 - m^2 \bigg ) \
- \ {\hbar \over 8 \pi} \bigg ( 3\lambda \phi^2 - m^2 \bigg )
\ \ln \bigg ( {3\lambda\phi^2 - m^2 \over 2m^2} \bigg ) \ .\eqn\EFFPOTONE$$
{}From here we can calculate the quantum corrections to the kink energy,
$$ \ephi^{potential} = \int dx \ V_1^{\delta\phi} \left(\phik(x)\right) \ ,
\eqn\EONEPHIPOT$$
where $\phik$ is the zero order (tree-level) kink solution,
eq.~\PKINK.\ The total kink energy is then
$$\eqalign{
E_{\rm kink}^{potential} \ &= \ {2\sqrt{2}\over 3} {m^3 \over \lambda} \
- \ \hbar m \ {3 \sqrt{2} \over 4 \pi } \cr
&- \ \hbar m {\sqrt{2} \over 8\pi}
\ \int_{-\infty}^{\infty} dz \bigg (
3\tanh^2 z - 1 \bigg ) \ \ln \bigg \vert {3 \tanh ^2 z
- 1 \over 2 } \bigg \vert \ + \ {\cal O}(\hbar^2) \cr
&= \ {2\sqrt{2}\over 3} \ {m^3 \over \lambda} \ \bigg [ 1 \ - \
0.3165 \ {\hbar \lambda \over m^2} \ + \ {\cal O}(\hbar^2) \bigg ]\  . \cr}
\eqn\ENERPOT$$
How ``effective" is the effective potential technique?
By comparing $E_{\rm kink}^{action}$ with
$E_{\rm kink}^{potential}$, eqs.~ \ENERACTION\
and \ENERPOT,\ we see that there is a difference in the quantum corrections
of about 50 percent. This is a significant discrepancy and clearly
demonstrates the insufficiency of the effective potential for inhomogeneous
configurations.

\chapter{Coupled fields}

In section~3 we focused on a single scalar field $\phi$ and computed the
quantum
fluctuations about its own non-homogeneous configuration. As there
exists only a
single coupling constant, $\lambda$, this system is rather restrictive;
$\lambda$ (together with {\it m}) fixes the kink, but at the same time
controls the coupling of the fluctuations to the background. As a result, the
Schr\"odinger equation~\SCHR\ turns out to be $\lambda$-independent. (By
rescaling the coordinate {\it x}, the dependence on {\it m} becomes trivial.)
There is, therefore, not much of an interesting parameter space to cover.
But there is a more basic problem with this simplified model. A
non-interacting scalar field may not adequately reflect the physical situation.
As we are eventually interested in the tunneling rate within, {\it e.g.}, the
electroweak phase transition, we have to treat the Higgs field with all its
interactions.

In the present section we challenge the effective
potential in a slightly more physical framework and couple $\phi$ to a second
scalar field $\chi$ (still confined to (1+1) dimensions). Although this falls
short of the real situation, in which the Higgs field is coupled to gauge
bosons and quarks in (3+1) dimensions, we hope that it will give us some hint
about the sufficiency or insufficiency of the effective potential.
In sections~4.1 and 4.2 we will again compute the total kink energy to
one-loop in the same two approaches discussed in section~3, and compare the
results. We will see how the introduction of a second scalar field allows us to
vary the coupling of the $\chi$-fluctuations to the
underlying soliton and therefore to find a regime where derivative
terms are indeed negligible.


There is also a regime in which derivative terms
are important. In section~4.3 we introduce a different approximation,
the Born approximation, for use in this regime. We recalculate
the kink energy within the Born approximation and compare
it with the accurate result of section~4.1.

\section{Effective Action for two fields}

Let us focus on the Lagrangian
$$ {\cal L} \ = \ {1\over 2} \partial_\mu\phi \partial^\mu\phi
+{1\over 2} \partial_\mu\chi \partial^\mu\chi
        - V_0 (\phi,\chi) \ ;\ \ \
V_0 \ = \  {\lambda\over 4} \bigg (\phi^2 - {m_0^2 \over \lambda}\bigg)^2 +
{1 \over 2} \ g \ \phi^2 \chi^2 \ . \eqn\LAGTWO$$
This model again permits a kink configuration $\phi_{\rm kink}$, eq.~\PKINK,\
along with $\chi \equiv 0$. The classical energy of the kink is unchanged from
eq.~\ECLKINK\ and is {\it g-independent}.
Now consider the quantum fluctuations about the kink. To
one-loop, the $\phi$ and $\chi$ fluctuations are independent, and therefore
yield two separate contributions to the energy shift.
The contributions from the $\phi$ field were calculated in the previous
section, here we concentrate on the $\chi$-fluctuations. Their
contribution to the kink energy again amounts to half the sum of the
eigenfrequencies $\omega_n^\chi$ of a Schr\"odinger equation, in this
case of
$$\bigg [ \ - {d^2 \over dx^2} + {g \over \lambda} m^2 \ {\rm tanh}
^2 \bigg({m x \over \sqrt {2}}\bigg ) \
\bigg ] \psi_n^\chi (x) \ = \ \big (\omega _n^\chi \big ) ^2
\psi_n^\chi (x) \ . \eqn\SCHRCHI$$
The coupling constants do not cancel out, but appear in the
combination $g /\lambda$ and can, in principle, attain {\it any} value within
our perturbation  expansion. (Note that our expansion parameters are
$\hbar \lambda /m^2$ and $\hbar g /m^2$.)
At times it will prove more convenient to use the dimensionless variables,
$\epsilon_n \equiv 2\ (\omega_n^\chi)^2/m^2$ and $z\equiv m x /\sqrt{2}$.
In terms of these variables,
eq.~\SCHRCHI\ becomes
$$\bigg [\ {d^2 \over dz^2} \ + \
\epsilon_n \ -\ 2 \ \gl \ {\rm tanh}^2 z
\ \bigg ] \ \psi_n^\chi (z) \ = \ 0 \ . \eqn\SCHRCHITWO$$

First let us consider the bound states of Eq. \SCHRCHITWO.
The ``potential'' in this
Schr\"odinger equation is $2(g/\lambda)\tanh^2z$,
its strength governed by the dimensionless ratio $g/\lambda$. Large
$g/\lambda$ corresponds to deep potentials; we expect these to have
many bound states. Small values of $g/\lambda$ should have only one
bound state.\footnote{\ddagger}{Recall the peculiarity of one
spatial dimension, in which
there always exists at least one bound state, no matter how
`shallow' the potential.} In fact, as shown in ref.~[\MORSE],
eq.~\SCHRCHITWO\ has {\it N} bound states with energies
$$\epsilon_n = 2 \ \gl - \bigg ( \gamma - n - {1 \over 2} \bigg )
^2 \ ;\ \ n=0,1,...,N-1<\gamma-1/2 \ , \eqn\EPSILON $$
where
$$\gamma\equiv \sqrt{2g/\lambda+1/4}. \eqn\EDEFGAMMA $$

Next we find the continuous eigenmodes of eq.~\SCHRCHITWO.\
These have
frequencies equal to those in the vacuum: $\omega_{\rm cont}^\chi =
\sqrt{k^2 + (g/\lambda)m^2}$. Corresponding to each frequency
are two states with definite (and opposite) values of parity and
generally also different phase shifts.
(Since the potential is invariant under parity, we
can use a basis in which all states have definite parity.) As we saw
in the previous section, it is necessary to find the phase
shift in order to sum up the energy of the continuous modes. To this
end, we note that
the problem at hand is equivalent to a scattering problem. Assuming an
incident wave from $z=-\infty$, the asymptotic form of the scattered
wave is [\MORSE]
$$ \psi^\chi \ \rightarrow \ {\cal N} \ \cases{ a e^{iqz}+b e^{-iqz}
& ; \ \ z \rightarrow -\infty \cr
e^{iqz} & ; \ \ z \rightarrow \infty  \cr } \eqn\WAVES $$
where
$$a \ = \ {\Gamma (1-iq) \ \Gamma (-iq)
\over \Gamma ({1 \over 2}+\gamma-iq)
\ \Gamma ( {1 \over 2} -\gamma -iq)} \ \ , \ \ \
b\ = \ -i \ {\cos ( \pi \gamma ) \over \sinh ( \pi q ) }
\ . \eqn\AB$$
${\cal N}$ is an irrelevant (complex) normalization factor and
$q=\sqrt{2} k / m$ is the dimensionless wavenumber.
In Appendix~B we derive the phase shifts for the two parity eigenstates
(the symmetric and antisymmetric parts of the scattering wave),
$$\delta_\chi^\pm \ = \ i \ \ln \bigg ( {a \over 1 \pm b} \bigg ) \ , \
\ {\rm and \ hence}
\ \ \delta_\chi \ \equiv \ \delta_\chi ^+ + \delta_\chi ^- \ = \
2 i \ \ln \bigg ( { a \over \vert a \vert } \bigg ) \ . \eqn\SHIFTS$$
For $\gamma$ a half-integer, $b=0$, implying that
the even and odd parity phase shifts are identical:
$\delta_\chi^+ = \delta_\chi^-$. (In terms of scattering theory, the
potential is reflectionless.)
This was the case for the single scalar field of section~3 (there
$\gamma=5/2$), and it was, therefore, unnecessary to consider
even/odd parity states separately.

In fig.~\FSHIFT\ we show the phase shifts as a function of
$q$ for fixed $g/\lambda$. The behavior of the
phase shift at low momenta is of particular interest and  we
therefore plot in fig.~\FSHIFTZERO\
the phase shifts at zero wavenumber
as a function of $g/\lambda$. $\dplus(0)$ ($\delta^- (0)$) changes
discontinuously as $\gamma$ passes through
$2.5,4.5,6.5,\ldots$ ($1.5,3.5,5.5,\ldots$). To be specific,
let us focus on the regime around $\gamma = 2.5$,
when $\dplus(0)$ changes
from $\pi$ to $3\pi$. According to eq.~\EPSILON,\ at this
$\gamma$ a new bound state appears. In fact each time a new bound
state appears, $\dplus(0)$ or $\dminus(0)$ changes by $2\pi$.
The connection between the phase shift and the number of bound
states is, of course, no accident, but a manifestation
of Levinson's theorem
[\LEVINSON,\BERG].\footnote{\natural}{We can count the
number of bound states by means of Levinson's
theorem for one-\hfill \break
dimensional potential scattering. The number of even
and odd states amounts to
$$N_\pm = \ {1 \over 2 \pi } \ \delta_\chi^\pm (0)
\mp \ {1 \over 4} \ \bigg ( \
\exp[i \delta_\chi^\pm (0)] - 1 \ \bigg ) \ . $$
For a clear discussion of Levinson's theorem in
quantum field theory, see ref.~\JACKIW.}

As in section~3, we quantize the continuous states in a box
and constrain {\it k} (separately for the even and odd
wavefunctions) by imposing periodic boundary conditions,
$$ k_n \ = \ {2\pi n\over L} \ - \ {\delta_\chi^\pm(k) \over L} \ .\eqn\KPM$$
Requiring $n\le n_{max}$ leads again to an ultraviolet cut-off,
$$ k \ \le \ \kmax \ \equiv \ \Lambda \ - \  {\delta_\chi^\pm(\kmax) \over
L} \ .\eqn\DKMAX $$
Because $\dplus(k) \simeq \dminus(k)$ for large {\it k}, the ultraviolet
cut-off for the symmetric and anti-symmetric states are identical.

In summing up the continuous eigenmodes we again replace the sum
by an integral, but this time separate out the even parity states from
the odd ones,
$$\eqalign{{\hbar \over 2} \sum \omega_{\rm cont}^\chi
&= \ {\hbar\over 2}
\int_{k_{\rm min}^+}^{\kmax} {dk\over 2\pi} \ \bigg ( L + {d\delta_\chi^+
\over dk} \bigg ) \ \omega_{\rm cont}^\chi \ + \ {\hbar\over 2}
\int_{k_{\rm min}^-}^{\kmax} {dk\over 2\pi} \ \bigg ( L + {d\delta_\chi^-
\over dk} \bigg ) \ \omega_{\rm cont}^\chi \ . } \eqn\SCONT$$
We have also introduced lower cutoffs $k_{\rm min}^\pm$, reflecting
the fact that below a minimum {\it n} all states have dropped into the
potential well and are counted as bound states. Due to the periodic
boundary conditions, we require
$$ k \ > \ k_{\rm min}^\pm \ \equiv \ {2 \pi n_{\rm min}^\pm \over L} \ - \
{\delta_\chi^\pm(k_{\rm min}) \over L} \ .\eqn\DKMIN $$
$n^\pm_{\rm min}$ can be identified with the
number of even and odd bound states,
$n^\pm_{\rm min}=n^\pm_{\rm discrete}$.
One might think that the lower limit in momentum is irrelevant as
it goes like $1/L$.
This is true except for the term in eq.~\SCONT\ which diverges with $L$.
Neglecting $k_{\rm min}^\pm$ in this term could lead
to a doublecounting of the discrete states.
It was merely a numerical coincidence that
in the one-field case of section~3 we could set $k_{\rm min}=0$. This is due
to the fact that for $\gamma$ half-integer $\delta_\chi$ is a multiple
of $2 \pi$, and, together with the exact number of bound states, leads
to $k_{\rm min}^\pm \equiv 0$.\footnote{\sharp}{In section~3
we have integrated from
$-k_{max}$ to $k_{max}$ (see eq.~\EONEPHI),\ which is the equivalent of
integrating the even and odd states independently from $0$ to $k_{max}$
(the integrand is symmetric).}
With the help of $k_{\rm min}$, eq.~\DKMIN,\ we can recast the sum over the
continuous eigenmodes,
$$\eqalign{{\hbar \over 2} \sum \omega_{\rm cont}^\chi
&= {\hbar \over 2} \int_0 ^{\kmax} {dk \over 2 \pi}
\bigg [ \ 2L + {d \over dk} \bigg ( \delta_\chi^+ + \delta_\chi^- \bigg )
\ \bigg ] \ \omega_{\rm cont}^\chi \cr
&\ \ \ \ \ \ - {\hbar \over 2} \ \bigg [ \
n_{\rm min}^+ + n_{\rm min}^- - {1 \over 2 \pi} \bigg
( \delta_\chi^+(0) + \delta_\chi^-(0) \bigg ) \ \bigg ] \
\omega_{\rm cont}^\chi (k=0) \cr
&= {\hbar \over 4 \pi } \int_0 ^{\kmax} dk
\ \bigg [ \ 2L \ \omega_{\rm cont}^\chi
- {k \over \omega_{\rm cont}^\chi }\ \delta_\chi(k) \ \bigg ] \cr
&\ \ \ \ \ \ + \ {\hbar \over 4 \pi} \ \delta_\chi (k_{max}) \
\omega_{\rm cont}^\chi (k_{max}) \
- \ {\hbar \over 2} \ n_{\rm min} \
\omega_{\rm cont}^\chi (0) \ , } \eqn\SCONTTWO$$
where we have made an integration by parts and defined
$ n_{\rm min} = n_{\rm min}^++n_{\rm min}^- $ (recall also that
$d \omega / dk = k /\omega$).
The last term above takes care of the previously mentioned
doublecounting: As the integral for the
continuum in eq.~\SCONTTWO\ extends over all states, we
have to subtract off the contributions from the bound states
($n_{\rm min}$ in total).

Finally note that the second term in the integral of eq.~\SCONTTWO\
is logarithmically
divergent. When we replace the bare mass with the physical one
in the classical energy, we induce a correction
$\hbar {\sqrt{2} m \over 4\pi } {g \over \lambda} \ln \bigg (
{4\Lambda^2 \over  m^2 g / \lambda } \bigg )$,
 which exactly cancels this divergence. (The same is true for
the term proportional to {\it L}, which gets canceled by the induced
``cosmological constant".)

Collecting all the contributions to the kink energy, {\it i.e.}, the
classical energy, the contributions from the renormalization, and
the shifts due to the quantum fluctuations of the $\phi$ and the $\chi$
fields about the inhomogeneous background,
we finally obtain (to order $\hbar$)
$$E_{\rm kink} ^{action} \ = \ E_0 \ + \ \ephi^{action}\ + \ \echi^{action}\
, \eqn\EKINKTWO$$
where
$$\eqalign{E_0 \ &= \ {2 \sqrt{2} \over 3} {m^3 \over \lambda} \ , \cr
\ephi^{action}\ &= \ \hbar m \ \bigg ( {1 \over 2\sqrt{6}}
- {3\over \sqrt{2} \pi} \bigg ) \ , \cr
\echi^{action}\ &= \
{\hbar \over 2} \sum \omega_{\rm discrete}^\chi
\ \ \ - \ {\hbar \over 2} \ n_{\rm discrete} \sqrt {g/\lambda} \ m
\cr
&\ \ \ - {\hbar \over 4 \pi }
\int _0 ^\Lambda dk \
{k \delta_\chi (k) \over \sqrt{k^2 + m^2 g/\lambda } }
\ + \ {\sqrt{2} \over 4 \pi} \hbar m \gl \ \ln
{4 \Lambda ^2\over m^2 g/\lambda} . } \eqn\EE$$
In $E_{\delta \chi}^{action}$ we have set
$n_{\rm min} = n_{\rm discrete}$.\footnote{\dagger}{The integral in the
$\chi$-correction is logarithmically
divergent; the divergence is cancelled by by the $\ln \Lambda$ term
in eq. \EE. Numerically, we can evaluate the sum of these two terms by
adding and subtracting the term
$-\int {dk \over k + c} = -\ln ({\Lambda +c \over c})$,
{\it c} an arbitrary positive constant.}
Fig.~4 shows the
energy shift due to $\chi$ fluctuations, $\echi^{action}$, as a function of
the dimensionless ratio of the couplings, $g/\lambda$.

\section{Effective potential method for two fields}

Let us now compare the above energy shift, induced by the $\chi$ field,
with the energy within the effective potential approach.
We again separate out the one-loop corrections in the effective potential from
the classical potential,
$$V_{\rm eff} \ = \ V_0 (m,\lambda) + V_1^{\delta\phi}(m,\lambda) +
V_1^{\delta\chi} (m,\lambda,g) \ . \eqn\EFFPOTTWO$$
The energy shift due to $V_1^{\delta\phi}$ has already been calculated in
section~3.2. Here we focus on the $\chi$ contribution (see eq.~(A9)),
$$V_1^{\delta\chi} (\phi) \ = \ {\hbar g \over 8 \pi} \ \bigg ( \
\phi ^2 - {m ^2 \over \lambda} \ - \
\phi^2 \ \ln {g \phi ^2 \over m^2 g/\lambda } \ \bigg )
\ . \eqn\POTONECHI$$
In complete analogy to eq.~\EONEPHIPOT,\ the energy shift is now given
by
$$\eqalign{\echi^{potential} &= \int_{-\infty}^\infty dx \
V_1^{\delta\chi} (\phi_{\rm kink}(x)) \cr
&= \ {\sqrt{2} \over 8 \pi } {g \over \lambda} \hbar m
\int_{-\infty}^\infty dz \bigg [
\tanh^2 z \ - \ 1 \ - \ \tanh^2 z \ \ln \bigg ( \tanh^2 z \bigg ) \bigg ]
\cr
&= \ {\sqrt{2} \over 8 \pi } \ {g \over \lambda} \hbar m
\ \bigg ( \ {\pi ^2 \over 2} - 6 \ \bigg ) \ \simeq \ -0.0599 \ \hbar m
\ {g \over \lambda} \ , } \eqn\ECHIEFFPOT$$
where we have used $\phi_{\rm kink}$ from eq.~\PKINK.\
By adding these $\chi$-corrections to the $\phi$-terms,
eq.~\ENERPOT,\ we obtain the total kink energy within the effective
potential approach,
$$\eqalign{E_{\rm kink}^{potential} \ &= \ E_{cl} \ + \
\ephi^{potential} \ + \ \echi^{potential} \cr
&\simeq \ {2 \sqrt{2} \over 3} {m^3 \over \lambda} \ \bigg [ \ 1 \ - \
0.3165\ {\hbar \lambda \over m^2 } \ - \ \ 0.0636 \ {\hbar g \over m^2}
 \ \bigg ] \ . } \eqn\EKINKPOT$$

In fig.~\FEN\ we compare $\echi^{action}$ with $\echi^{potential}$
for varying $g/\lambda$. For small $g/\lambda$,
$\echi^{potential}$ differs considerably from $\echi^{action}$
and the effective potential is only a poor approximation
to the effective action. However, for $g/\lambda \rightarrow \infty$,
$\echi^{action}$ asymptotically approaches $\echi^{potential}$.
Qualitatively we can understand this behavior in the following way:
The $\chi$-field has a mass $m_\chi \simeq m \sqrt{g/\lambda}$
and hence varies typically on a scale $m_\chi^{-1}$. On the other hand,
the kink extends over a region $m^{-1}$. If
$m_\chi^{-1} \ll m^{-1}$ ($g/\lambda \gg 1$) then the $\chi$-field
oscillates vehemently along the kink profile and is not affected
by the background variation. Here the effective potential
is valid. In the opposite limit
($g/\lambda \ll 1$), the $\chi$-field changes at least as
slowly as the kink and therefore probes the underlying configuration.
In this case derivative terms are non-negligible.

Another way to understand the fact that $E_{\delta\chi}^{potential}$
is a good approximation when $g/\lambda$ is large is to compute
the relative contributions of the next terms in the
derivative expansion. Let us consider the Feynman diagram in
fig.~\FLOOP. Its contribution to the effective action is weighted
by the Feynman integral, which can be expanded in powers of the
external momentum $p=p_1+p_2$:
$$\eqalign{
 \int {d^2k\over (k^2-\mc)( (p+k)^2 - \mc)}
&= \int {d^2k\over (k^2-\mc)^2}
\cr &+~\int {d^2k\over (k^2-\mc)^3} \left( p^2 + {4(p\cdot k)^2\over
k^2 - m_\chi^2} \right) + \ldots } \eqn\FEYNMAN $$
The first term on the right hand side is of order $1/m_\chi^2$,
while the second $\sim p^2/m_\chi^4$.
Since the kink
field varies on spatial scales of about $m^{-1}$, the factors of $p$
are ${\cal O}(m)$, so that the second integral
is of order $m^2/m_\chi^4$. Therefore, the first derivative correction
is
a factor of $m^2/\mc = \lambda/g$ smaller (or larger) than the
non-derivative terms in the effective potential. The derivative expansion
is therefore an expansion in the ratio of Compton wavelengths of the
background field and the fluctuating field. This explains
why in the one field case analyzed in section~3, the corrections from
derivative
terms are significant: the expansion parameter is of order one.

The derivative terms are down by at least ${\cal O} (\lambda /g)$ and can
therefore be neglected for $g/\lambda \rightarrow \infty$. As already
pointed out, in the other limit they are important and the full
effective action has to be evaluated. As we will demonstrate now, it is
precisely in this regime where one can rely on additional
approximations and thereby often simplify the computations.

\section{Born approximation}

We confined ourselves to a kink in (1+1) dimensions due to its simplicity.
Most realistic models do not allow for a closed solution (within
perturbation theory) and one has to rely on approximations.
The problem usually pins down to solving the Schr\"odinger equation.
In the worst case, the eigenvalues of the discrete levels have to be found
numerically (after all, there are only a finite number of bound states).
It may, however, not be that simple to sum up the eigenfrequencies of the
continuum and one would like to have a way of approximating their contribution.
Recall that to sum up the continuum eigenfrequencies, one need only know
the phase shift (the frequencies themselves are trivial).
Fortunately, there are circumstances where one can approximate the phase shift
without exactly solving the
Schr\"odinger equation. This is especially true when the fluctuations are
only weakly coupled to the underlying kink, or in other words,
the ``potential" in the Schr\"odinger equation is ``shallow".
In our model, this corresponds to $g/\lambda \ll 1$.
In this limit we can use the Born approximation.

Let us rewrite eq.~\SCHRCHITWO. Since for continuum states, $\epsilon_n
= q^2 + 2g/\lambda$, we find
$$\bigg [\ {d^2 \over dz^2} \ + \ q^2 \ + \ u(z) \
\bigg ] \ \psi_n^\chi (z) \ = \ 0 \ , \ \ {\rm where} \ \
u(z)={2 \over \cosh ^2 z} \ {g \over \lambda} \ ; \eqn\SCHRBORN$$
$u(z)$ is the relevant ``potential.''

Next we expand the wavefunction around the unperturbed wave
$\psi_0 (z) = \exp (iqz)$,
incident from the left (we take ${\cal N} = 1/a$ in
eq.~\WAVES).\ The total wavefunction then reads
$$\psi^\chi (z) \ = \ \psi_0 (z) + \psi _1 (z) \ . \eqn\PSIBORN$$
For perturbative ``potentials" ($g/\lambda \ll 1$), $\psi_1$ can be evaluated
within the Born approximation,
$$\psi_1 (z) \ = \ - \int_{-\infty} ^\infty dy \ G(z,y) \ u(y) \
\psi_0 (y) \ , \eqn\PSIBORN$$
where $G(z,y)$ denotes the one-dimensional Green's function,
$$G(z,y) \ = \ - {i \over 2 q} \ e^{iq \vert z-y \vert} \ . \eqn\GREEN$$
Specifically, for $z \rightarrow \infty$ we obtain
$$\psi_1 (z \rightarrow \infty) \ = \ {i \over 2 q} \ e^{iqz} \int _{-\infty}
^\infty u(y) dy \ = \ {2i \over q} \gl \ e^{iqz} \ . \eqn\PSIINF$$
By adding this to the unperturbed wave and comparing to eq.~\WAVES,\ we
can fix {\it a},
$$a \ = \ \bigg ( \ 1 + {2i \over q} \gl \ \bigg )^{-1} \ , \eqn\ABORN$$
and hence the Born approximated phase shift,
$$\eqalign{ \delta_\chi ^{Born} \ &= \ 2i \ \ln \bigg (
{a \over \vert a \vert} \bigg )
\ = \ i \ln \bigg ( {a \over a^\ast} \bigg ) \cr
&= \ -i \ \ln \bigg ( {1 + i {2 g/\lambda \over q} \over
1 - i {2 g/\lambda \over q} } \bigg ) \ = \ 2 \tan ^{-1} \bigg (
{\sqrt{2} m g/\lambda \over k} \bigg ) \ . } \eqn\DELTABORN$$
With the help of $\delta_\chi^{Born}$ we can calculate the contribution from
the continuous eigenfrequencies, and specifically the integral
$$\eqalign{{\hbar \over 4 \pi} \int _0 ^{k_{max}} dk \
{d \delta_\chi ^{Born}
\over dk} \ \omega_{\rm cont}^\chi \ &= \ - {\hbar m \over 4 \pi} \ \bigg [
\ \sqrt{2} \ \gl \ \ln {4 \Lambda^2 \over m^2 g/\lambda} \cr
&\ \ \ + \sqrt{ \gl \bigg ( 1-2\gl \bigg )} \bigg
( \sin^{-1} \bigg (1 - 4\gl \bigg ) + {\pi \over 2}
\bigg ) \ \bigg ] \ . } \eqn\CONTBORN$$
(This equation only holds in its present form for $g/\lambda < 1/2$.)
For the (single) discrete eigenfrequency we will use the exact result
according to eq.~\EPSILON.\
(In most cases one will have to evaluate the discrete eigenvalue(s),
if there are any, numerically.\footnote{\ddagger}{In more than one spatial
dimension there are likely to be no discrete states for shallow
potentials.}) After collecting also the contributions from the
renormalization, the total energy shift due to the $\chi$ field adds up
to
$$\eqalign{E_{\delta \chi}^{Born} \
&= \ {\hbar m \over 2 \sqrt{2}} \ \bigg [ \ 2 \ {g \over \lambda} -
\bigg ( \sqrt {2 \gl + {1 \over 4}} - {1 \over 2} \bigg ) ^2 \ \bigg ]
^{1/2} \ - \ \hbar m \ {\sqrt{2} \over 2 \pi} \ \gl \cr
&\ \ \ - {\hbar m \over 4 \pi} \sqrt{ \gl \bigg ( 1 - 2 \gl \bigg ) }
\bigg ( \sin ^{-1} \bigg ( 1 - 4 \gl \bigg ) + {\pi \over 2}
\bigg ) \ - \ {\hbar m \over 4} \sqrt{ \gl } \ . } \eqn\ENERBORN $$
The last term again prevents us from doublecounting the bound state.

In fig.~\FEN\ we plot the energy shift from the Born approximation and
compare it
with the previous results. The Born approximation does rather well for very
small $g/\lambda$, but starts to deviate significantly for
$g/\lambda \gsim 0.01$. Note, however, that the effective potential approach is
always worse than the Born approximation in the small $g/\lambda$ regime.

\section{Conclusions}

One of the prerequisites of understanding a first order phase transition
is the calculation of the bubble nucleation rate, or the decay rate of
the false vacuum. This involves, by definition, inhomogeneous
configurations. In evaluating the decay rate, one has to consider quantum
fluctuations. The question then arises, whether it is adequate to drop
the derivative terms in the effective action and only retain the
effective potential, or whether these derivative corrections are
important. In other words, do we have to integrate out quantum
fluctuations about the {\it inhomogeneous} configuration, or is it
sufficient to consider the fluctuations about a {\it constant} background?
In the attempt to answer this question, we confined ourselves to a simple
model, a kink in (1+1) dimensions and zero temperature.
We first computed the energy shift from quantum fluctuations
about the inhomogeneous background and then compared it to the energy
from the effective potential. The discrepancy in the two results clearly
demonstrates the insufficiency of the effective potential. Note that
although quantum effects are small by definition, it is precisely these
corrections, which may fix the order of a given phase transition. Any
error may therefore obscure the real situation. We next coupled the kink
field to a second scalar field and also included its quantum fluctuations in
the energy calculation. The addition of the second field allowed us to
vary the mass of the fluctuations and their coupling to the kink and
thereby to span a larger parameter space. In the limit where the
mass of the additional fluctuations is large, the fluctuations do not
probe the kink profile efficiently. This is the regime, where the
effective potential represents a valid approximation to the full action.
In the limit of small mass, however, the derivative terms
are important and one cannot rely on the effective potential.
It is interesting to note that it is precisely in this regime, where one
can resort to additional approximations, as for example the Born
approximation. In more realistic
situations it may be impossible to derive an exact result and one is
forced to rely on such approximations. We therefore recalculated the
energy shift within the Born approximation and compared it to the exact
result. Although, in our case,
the approximated energy falls somewhat short of the exact result, its
estimate is significantly more accurate than the effective potential
result.

In conclusion we should point out, that the current calculations have been
carried out in a very simple system and their results may not necessarily
be extrapolated to the analysis of a first order phase transition in,
{\it e.g.}, the electroweak theory. Higher dimensions, finite
temperature effects and the introduction of fermions and gauge fields
can possibly change the conclusions. In light of our results, it must,
however, be kept in mind, that there is no {\it a priori} justification to
neglect derivative terms, but that the validity of the effective
potential for inhomogeneous configurations must be checked from case to
case.

\bk
\noindent
ACKNOWLEDGEMENTS

It is a pleasure to thank Erick Weinberg for many helpful discussions about
effective potentials and derivative terms. We also acknowledge valuable
discussions with Robert Brandenberger, Brian Greene,
Chris Hill, Dallas Kennedy, and Esteban Roulet.
This work was supported in part by the DOE and NASA grant
NAGW-2381 at Fermilab. B.G. also acknowledges DOE grant
\#DE-FG03-91ER (40662 Task C) at UCLA.
\bk

\centerline{APPENDIX A}

In this Appendix we outline our renormalization scheme for the two field case.
We start with the tree level potential of eq.~\LAGTWO,\
$$V_0 \ = \  {\lambda\over 4} \bigg (\phi^2 - {m_0^2 \over \lambda}\bigg)^2 +
{1 \over 2} g \ \phi^2 \chi^2 \ . \eqno(A1) $$
Following standard procedure, the effective potential to one-loop becomes
$$\eqalign{V_{\rm eff} \ &= \ V_0 \ + \
{\hbar \over 8 \pi} \bigg ( 3 \lambda \phi^2 -
m_0^2 \bigg ) \bigg [ \ 1 + \ln {4 \Lambda^2 \over 3 \lambda \phi^2 - m_0^2}
\ \bigg ] \cr &\ \ \ \ \ + \ {\hbar \over 8 \pi}
g \phi^2 \bigg [ \ 1 + \ln {4 \Lambda^2 \over g
\phi^2 } \ \bigg ] \ + \ 2 \ {\hbar \Lambda^2 \over 4 \pi} \ , }
\eqno(A2) $$
with minimum at $\phim$,
$$\phi_M^2 \ = \ {m_0^2 \over \lambda} - {\hbar \over 4 \pi}
\ \bigg [ \ 3 \ln {2 \Lambda^2 \over m_0^2 } + {g \over \lambda} \ln
{4 \Lambda^2 \over m_0^2 g/\lambda} \ \bigg ] \ + \ {\cal O} (\hbar ^2 ) \ .
\eqno(A3) $$
The regularization parameter $\Lambda$ is the ultraviolet momentum
cutoff.
(We can trivially retrieve the effective potential
for the single scalar field model of section~3 by setting $g=0$
and neglecting one of the terms $\hbar \Lambda^2/(4 \pi)$ in eq.~(A2).)

We define the physical (renormalized) parameters by ``minimal
subtraction", which `pins down' to the identification of
the bar coupling constant $\lambda$ and the physical one,
$$\lambda_{phys} \ \equiv \ \lambda \ . \eqno(A4)$$
Note that this identity holds to all orders in perturbation theory.
(As the $\chi$-field has vanishing expectation value and we are only
working to one-loop, we do not have do worry about the renormalization
of {\it g}.) We define the renormalized mass by
$$\eqalign{{1 \over 2} \ {d^2 \veff \over d\phi^2}(\phi = \phim ) \ &= \
m_0 ^2 \ - \ \lambda \ {\hbar \over 4 \pi } \ \bigg ( \  {9 \over 2} +
{g \over \lambda} \ \bigg ) \cr
&\ \ \  \ - \ \lambda {\hbar \over 4 \pi} \ \bigg ( \ 3 \ln
{2 \Lambda^2 \over m^2} + {g \over \lambda} \ \ln {4 \Lambda \over
m^2 g/\lambda} \ \bigg ) \cr
&\equiv \ m^2 \ - \ \lambda \ {\hbar \over 4 \pi }
\ \bigg ( \  {9 \over 2} + {g \over
\lambda} \ \bigg ) \ ,} \eqno(A5)$$
where {\it m} without the subscript denotes the physical mass. From
here,
$$m_0^2 \ = \ m^2 \ + \ \lambda \ {\hbar \over 4 \pi} \ \bigg ( \ 3 \ln
{2 \Lambda^2 \over m^2} \ + \ {g \over \lambda} \ \ln {4 \Lambda \over
m^2 g/\lambda} \ \bigg ) \ . \eqno(A6)$$
The virtue of this renormalization scheme lays in the absence of any
{\it finite} quantum corrections to the mass. Consequently, the tree-level
energy of the kink, if written in terms of the physical
parameters, contains only infinite corrections and no
additional finite terms.

The requirement that $V_{\rm eff} (\phim) = 0$ generates a
vacuum energy (``cosmological constant"),
$$\eqalign{\Gamma \ &\equiv \ \Gamma^\phi \ + \ \Gamma^\chi \cr
&= \ - {\hbar \over 4 \pi} \ \bigg [ \ m^2 \bigg ( 1 + \ln
{2 \Lambda^2 \over m^2 } \bigg ) \ + \ \Lambda^2 \ \bigg ] \cr
&\ \ \ - \ {\hbar \over 4 \pi} \ \bigg [ \
{m^2 \over 2 } \gl \bigg ( 1 + \ln {4 \Lambda ^2 \over m^2 g /
\lambda } \bigg ) \ + \ \Lambda^2 \ \bigg ] \
+ \ {\cal O} (\hbar^2) \ . } \eqno(A7) $$
(For the single scalar field case we omit the whole second term.)

Inserting (A6) and (A7) into eq.~(A2) yields the effective potential in terms
of
the renormalized parameters,
$$V_{\rm eff} \ = \ V_0(m,\lambda) \ + \ V_1 ^{\delta \phi} (m,\lambda) \
+ \ V_1 ^{\delta \chi} (m,\lambda,g) \ , \eqno(A8)$$
where $V_0$ is the tree-level potential (A1) in terms of the physical
mass, and
$$\eqalign{V_1^{\delta \phi} \ &= \
{3 \hbar \over 8 \pi} \ \bigg ( \ \lambda \phi^2 - m^2 \ \bigg ) \
- \ {\hbar \over 8 \pi} \ \bigg (3\lambda \phi^2 - m^2
\bigg ) \ \ln \bigg ({3\lambda\phi^2 - m^2\over 2m^2} \bigg ) \ , \cr
V_1^{\delta \chi} \ &= \ {\hbar g \over 8 \pi} \ \bigg ( \ \phi^2 - {m^2 \over
\lambda} - \phi^2 \ \ln {g \phi^2 \over m^2 g /\lambda } \
\bigg ) \ . } \eqno(A9)$$
\bk

\centerline{APPENDIX B}

In this Appendix we evaluate the phase shifts,
associated with $\chi$-waves scattering off the kink.

The ``potential" in the Schr\"odinger equation \SCHRCHI\
is symmetric under the reflection
$x \rightarrow -x$. This allows us to write the wavefunction
$\psi$ as a linear combination of parity eigenstates, {\it i.e.}, in terms
of symmetric and antisymmetric functions,
$$\psi^\chi (x) \ = \ \psi_+ (x) \ + \ \psi_- (x) \ \ \ {\rm with} \ \ \ \
\psi_\pm(-x) = \pm \psi_\pm (x) \ . \eqno(B1) $$
For simplicity, let us define 1-dimensional ``polar" coordinates
$r=\vert x \vert$ and $\mu = {\rm sgn} (x)$ and recast the symmetric
and antisymmetric waves as~[\BUTKOV]
$$\psi_+ \ = \ D_+ \ e^{ikr} \ + \ C_+ \ e^{-ikr} \ \ , \ \ \
\psi_- \ = \ \mu \ \bigg ( D_- \ e^{ikr} \ + \ C_- \ e^{-ikr} \bigg ) \ ,
\eqno(B2) $$
where the coefficients are given in terms of {\it a} and {\it b} of eq.~\AB,\
$$D_\pm \ = \ { {\cal N} \over 2 } \ \bigg ( \ 1 \pm b \ \bigg ) \ \ , \ \ \
C_\pm \ = \ \pm { {\cal N} \over 2 } \ a \ . \eqno(B3) $$
The inflection invariance of the potential implies~[\BUTKOV] that
$$\vert D_+ \vert ^2 \ = \ \vert C_+ \vert ^2 \ \ {\rm and} \ \ \
\vert D_- \vert ^2 \ = \ \vert C_- \vert ^2 \ , \eqno(B4) $$
and hence $D_+$ ($D_-$) can at most differ from $C_+$ ($C_-$)
by a phase,
$$D_+ \ = \ C_+ \ e^{i\delta_\chi^+} \ \ , \ \ \
D_- \ = \ C_- \ e^{i\delta_\chi^-} \ . \eqno(B5) $$
This, together with eqs.~(B3), yields
$$\delta_\chi^\pm \ = \ i \ \ln \bigg ( {a \over 1 \pm b} \bigg ) \ \ \
{\rm and} \ \ \ \delta_\chi \ \equiv \ \delta_\chi ^+ + \delta_\chi ^- \ = \
2 i \ \ln \bigg ( { a \over \vert a \vert } \bigg ) \ . \eqno(B6) $$

\vfill\eject
\refout

\vfill\eject
\figout

\bye